\documentclass{bowen}
\usepackage{graphicx}
\usepackage{amsmath}
\usepackage{latexsym}
\usepackage{graphicx}
\usepackage{subfig}
\usepackage{times}
\usepackage{authblk}
\usepackage{epstopdf}
\usepackage{multirow}
\usepackage{color}
\usepackage{latexsym}
\usepackage{amssymb}
\usepackage{eufrak}
\usepackage{euscript}
\usepackage[latin1]{inputenc}
\usepackage{pstricks,pst-coil}
\usepackage{animate}
\usepackage{pst-plot,pst-xkey}

\titleformat{\section}
 {\normalfont\normalsize\mdseries\centering}{\thesection}{1em}{}

\titleformat{\subsection}
{\normalfont\normalsize\mdseries\itshape}{\thesubsection}{1em}{}

\titleformat{\subsubsection}
{\normalfont\normalsize\mdseries\itshape}{\thesubsubsection}{1em}{}

\makeatletter
\let\sectionorig\section
\def\@sectionorig#1{\sectionorig*{\MakeUppercase{#1}}}
\def\@@sectionorig#1{\sectionorig{\MakeUppercase{#1}}}
\renewcommand{\section}{\@ifstar{\@sectionorig}{\@@sectionorig}}
\long\def\@makecaption#1#2{%
  \vskip\abovecaptionskip
  \sbox\@tempboxa{#1 #2}%
  \ifdim \wd\@tempboxa >\hsize
    #1 #2\par
  \else
    \global \@minipagefalse
    \hb@xt@\hsize{\hfil\box\@tempboxa\hfil}%
  \fi
  \vskip\belowcaptionskip}
\makeatother

\renewcommand*{\thesection}{\Roman{section}.}
\renewcommand*{\thesubsection}{\Alph{subsection}.}
\renewcommand*{\thesubsubsection}{\arabic{subsubsection})}

\title{\fontsize{24}{24} \selectfont A non-abelian model $SU(N) \times SU(N)$ \\ 
}

\author[1]{\fontsize{11}{11} \selectfont M. J. Neves}
\author[2]{\fontsize{11}{11} \selectfont R. Doria}
\affil[1]{\normalsize Departamento de F\'{i}sica,
Universidade Federal Rural do Rio de Janeiro, BR 465-07, 23890-971, Serop\'edica, RJ, Brazil}
\affil[2]{\normalsize AprendaNet Inform\'{a}tica, Petr\'{o}polis - RJ, Brazil}

\date{\vspace{-0.6cm}{\normalsize
$\mbox{}^1$mariojr@ufrrj.br;
$\mbox{}^2$doria@aprendanet.com.br
}}

\begin{document}
\maketitle
\thispagestyle{fancy}
\bibliographystyle{unsrt} 
%

\noindent

\textbf{Abstract} -
A composite non-abelian model $SU(N) \times SU(N)$ is proposed as possible
extension of the Yang-Mills symmetry. We obtain the corresponding gauge
symmetry of the model and the most general lagrangian invariant by $SU(N)
\times SU(N)$. The corresponding Feynman rules of the model are studied.
Propagators and vertices are derived in the momentum space. As physical
application, instead of considering the color symmetry $SU_{c}(3)$ for $QCD$%
, we substitute it by the combination $SU_{c}(3) \times SU_{c}(3)$. It
yields a possibility to go beyond $QCD$ symmetry in the sense that quarks
are preserved with three colors. This extension provides composite quarks in
triplets and sextets multiplets accomplished with the usual massless gluons
plus massive gluons. We present a power counting analysis that satisfies the
renormalization conditions as well as one studies the structure of radiative
corrections to one loop approximation. Unitary condition is verified at tree
level. Tachyons are avoided. For end, one extracts a BRST symmetry from
lagrangian and Slavnov-Taylor identities.
\\
\\
Keywords - \textit{Extended Yang-Mills Symmetry , Beyond Standard Model , Quantum Chromodynamics}.

\section{\textsc{Introduction}}

A non-Abelian model for composite fields is presented for investigating an
extension to the Yang-Mills case \cite{Yang54}. It yields the possibility to
explore an extended symmetry having contributions that go beyond to the
Yang-Mills symmetry \cite{DoriaHelayelMario}. For example, the possible
insertion of mass terms into the lagrangian no need breaking gauge symmetry
for non-abelian gauge fields.
The description of the interactions by means of composite fields has already
been considered by J. Schwinger \cite{Schwinger57}.
Also, others approaches by means of group composition were discussed for a
description of the Lepton-Hadron interaction beyond Weinberg-Salam-Glashow
electroweak model, for details \textit{see} \cite{PatiSalamPisano}. An
interesting search on description of massive non-abelian gauge fields is
given by use the Stueckelberg formalism as alternative to description of the
Standard Model, by including Higgs mechanism in breaking spontaneous
symmetry \cite{Ruegg03}.

The possibility to go beyond Yang-Mills symmetry is presented in this work
by considering the group $SU(N)$ as the combination $SU(N)\times SU(N)$.
Based on direct product we define this operation between two independent
non-abelian groups \cite{WybourneGilmoreLivro}. Notice that it is an
approach diverse from usual direct product in grand unification. Instead of
taking the Cartesian product of two groups $SU(N)\times SU(N)$, it
considers a common gauge group $SU(3)$ being rotated by two fundamental
representation, which means that we are just tensoring two fundamental
representations of the same $SU(N)$. Physically interpreting, we shall
propose a composite Quantum Chromodynamics $SU_{c}(3)\times SU_{c}(3)$
model. This means that instead of only $SU_{c}(3)$ as proposed by M.
Gell-Mann \cite{Gellman} and others, it realizes an extension to QCD in the
sense that preserves the experimental result that quarks contain three
colors. At this way introduces the possibility of having triplets and
sextets quarks and it yields the presence of massive gluons together with
the usual massless gluons case. 
The study of possible exotics Quantum Chromodynamics was discussed by \cite%
{Jaffe,NakanoBarminStepanyan} in which it can reveal the existence of exotic
barions in hadron spectroscopy.



The outline of the paper is organized as follows. The second section
introduces the symmetry gauge of $SU(N)\times SU(N)$ with non-abelian gauge
fields and quarks sectors, and shows the complete lagrangian invariant by
those symmetry transformations. In section 3 we begin a program for
renormalization of this model by analyzing the power counting and radiative
corrections to one-loop approximation. In section 4 one extracts a BRST
symmetry from effective quantum lagrangian and such symmetry leads to an
Slavnov-Taylor for $SU(N)\times SU(N)$. These quantities are necessaries to
full renormalization of the model. Finally, section 5 is left for the
concluding remarks on the prospected model $LHC$ possibilities.

\pagebreak

\section{\textsc{A non-abelian model for symmetry $SU(N)\times SU(N)$}}

\subsection{Gauge fields sector of $SU(N)\times SU(N)$}

Consider a fermionic matter field $\chi$ composite by the direct product
\cite{DoriaHelayelMario}
\begin{eqnarray}\label{quarkcomp}
\chi=\psi\otimes\phi \; ,
\end{eqnarray}
in which $\psi_{i}$ and $\phi_{i} \; (i=1,2,...,N)$ are independents spinors
and scalars fields both in the fundamental representation of each $SU(N)$ in
question, respectively. The fields $\left( \, \psi \, , \, \phi \, \right)$ have independent local
transformation
\begin{eqnarray}\label{transfglobais12}
\psi^{\prime}(x)=U_{1}(x) \, \psi(x) \hspace{0.5cm} \mbox{and} \hspace{0.5cm}
\phi^{\prime}(x)= U_{2}(x) \, \phi(x) \hspace{0.3cm} , \hspace{0.3cm} \mbox{with}
\hspace{0.3cm} U_{1}(x)=e^{\, i \, t_1^{a} \; \omega_{1}^{a}(x)} \hspace{0.3cm} %
\mbox{and} \hspace{0.3cm} U_{2}(x)=e^{ \, i \, t_2^{a} \; \omega_{2}^{a}(x)} \; ,
\end{eqnarray}
where $\left( \, t_{1}^{a} \, , \, t_{2}^{a} \, \right)$ are two independents generators of two
non-abelian groups $SU(N)$, satisfying the commutation relations
\begin{eqnarray}\label{relcomutacao12}
\left[ \, t_{1}^{a} \, , \, t_{1}^{b} \, \right]=i \, f^{abc} \, t_1^{c}
\hspace{0.5cm} \mbox{and} \hspace{0.5cm}
\left[ \, t_{2}^{a} \, , \, t_{2}^{b} \, \right]=i \, f^{abc} \, t_{2}^{c} \; ,
\hspace{0.3cm} \mbox{with}
\hspace{0.3cm} a=1,2,...,N^{2}-1 \; ,
\end{eqnarray}
and $\omega_{1}$ and $\omega_{2}$ are real functions. Notice that we have
considered the same structure constant of group $f^{abc}$ for the two
independents groups. Using properties of the direct products and the
transformations above, we obtain the local transformation $SU(N)\times
SU(N) $
\begin{eqnarray}\label{transflocaisU}
\chi^{\prime}(x)=U(x) \, \chi(x)
\hspace{0.5cm} , \hspace{0.5cm} U(x)=U_{1}(x) \otimes
U_{2}(x) \; .
\end{eqnarray}
Clearly, the spinor $\psi$ only belongs to $SU(N)$-left and scalar $\phi$
belongs to $SU(N)$-right of the product $SU(N)\times SU(N)$, and $\chi$ is
a fermion that belongs to full symmetry $SU(N)\times SU(N)$ whose
components are $\chi_{i} \; (i=1,2,...,N^{2})$ .

Nextly one shall be interested in establish a dynamic for these fermions $%
\chi$. For introducing the non-abelian gauge fields, a composite covariant
derivative based on representation product is proposed
\begin{eqnarray}\label{DAB}
D_{\mu}(A,B)=D_{\mu}(A)\otimes \mathbf{1} + \mathbf{1} \otimes D_{\mu}(B) \; ,
\end{eqnarray}
where each covariant derivative $D_{\mu}(A)$ and $D_{\mu}(B)$ act on $\psi$
and $\phi$, respectively
\begin{eqnarray}\label{DcovarianteAB}
D_{\mu}(A) \, \psi(x)=\left( \, \partial_{\mu}+i \, g_{1} \, A_{\mu} \, \right) \, \psi(x)
\hspace{0.5cm} \mbox{and} \hspace{0.5cm}
D_{\mu}(B) \, \phi(x)=\left( \, \partial_{\mu}+i \, g_{2} \, B_{\mu} \, \right) \, \phi(x) \; .
\end{eqnarray}
It is remarkable to notice that the covariant derivative of (\ref{DAB})
fulfills the requirement of satisfying the Jacobi identity.
This is why we can undertake that $D_{\mu}(A,B)$ is actually a covariant
derivative. Notice that we adopt here a different and alternative procedure
instead of grouping the different gauge potentials inside a single covariant
derivative. We propose a combined covariant derivative built up from a
different covariant derivative for each group factor.

The gauge fields $\left( \, A^{\mu} \, , \, B^{\mu} \, \right)$ transform in accord with
\begin{eqnarray}\label{transfcamposAB}
A_{\mu}^{\;\prime}=U_{1} \, A_{\mu} \, U_{1}^{-1}+\frac{i}{g_1} \,
\left(\partial_{\mu}U_{1} \right) \, U_{1}^{-1}
\hspace{0.5cm} \mbox{and} \hspace{0.5cm}
B_{\mu}^{\;\prime}=U_{2} \, B_{\mu} \, U_{2}^{-1}+\frac{i}{g_2} \,
\left(\partial_{\mu}U_{2}\right) \, U_{2}^{-1} \; ,
\end{eqnarray}
in which $\left( \, t_{1}^{a} \, , \, t_{2}^{a} \, \right)$ are basis of $\left( \, A^{\mu}\, , \, B^{\mu} \, \right)$
\begin{eqnarray}\label{BasesAB}
A_{\mu}=\sum_{a=1}^{N^2-1}A_{\mu}^{a} \, t_{1}^{a}
\hspace{0.5cm} \mbox{and} \hspace{0.5cm}
B_{\mu}=\sum_{a=1}^{N^2-1}B_{\mu}^{\;a} \, t_{2}^{a} \; ,
\end{eqnarray}
respectively. The constants coupling $g_{1}$ and $g_{2}$ are associated to
gauge fields $\left( \, A^{\mu} \, , \, B^{\mu} \, \right)$, respectively.
By using the definition (\ref{DAB}) and some properties of direct product one gets the following
transformation
\begin{eqnarray}\label{DcovTransf}
D_{\mu}(A,B)^{\prime}=U \, D_{\mu}(A,B) \, U^{-1} \; .
\end{eqnarray}
Therefore a symmetry $SU(N) \times SU(N)$ based on construction of direct
product is established for gauge fields $\left( \, A^{\mu} \, , \, B^{\mu} \, \right)$
and fermions $\chi$. If one carries on the fields $\left( \, A^{\mu} \, , \, B^{\mu} \, \right)$,
it will be possible to introduce mass terms by already known mechanisms, like Higgs mechanism
and Stueckelberg fields. However we intend to follow a different procedure.

As a next step we make the variable change
\begin{equation}\label{ABGC}
g_{1} \, A_{\mu }^{\;a}=g_{1} \, G_{\mu }^{\; a}+g_{2} \, C_{\mu }^{\; a}
\hspace{1cm} \mbox{and} \hspace{1cm}
g_{2} \, B_{\mu }^{\;a}=g_{1} \, G_{\mu }^{\; a}-g_{2} \, C_{\mu}^{\; a} \; ,
\end{equation}
in which $\left( \, G^{\mu} \, , \, C^{\mu} \, \right)$ will be the physical fields that we are
interested. Substituting (\ref{ABGC}) in (\ref{DAB}), it yields
\begin{equation}\label{DcovGC}
D_{\mu }(A,B) \, \chi(x) =D_{\mu}(G,C) \, \chi(x) =\left( \, \partial _{\mu }+i \, g_{1} \,
G_{\mu}+i \, g_{2} \, C_{\mu} \, \right) \, \chi(x) \; .
\end{equation}
where the new fields $\left( \, G^{\mu} \, , \, C^{\mu} \, \right)$ are Lie algebra valued on a new
basis of generators, $\{T^{a}\}$ and $\{t^{a}\}$, as
\begin{equation}\label{GTCt}
G_{\mu }=\sum_{a=1}^{N^{2}-1}G_{\mu }^{\;a} \, T^{a}
\hspace{0.5cm} \mbox{and} \hspace{0.5cm}
C_{\mu }=\sum_{a=1}^{N^{2}-1}C_{\mu }^{\;a} \, t^{a} \; ,
\end{equation}
where
\begin{equation}\label{Tt}
T^{a}=t_{1}^{a}\otimes \mathbf{1}+\mathbf{1}\otimes t_{2}^{a}\hspace{0.5cm}%
\mbox{and}\hspace{0.5cm}t^{a}=t_{1}^{a}\otimes \mathbf{1}-\mathbf{1}\otimes
t_{2}^{a}\;.
\end{equation}
Notice these new generators satisfy the commutation relation
\begin{equation}\label{RelComutTt}
\left[ \, T^{a} \, , \, T^{b} \, \right]=i \, f^{abc} \, T^{c}
\hspace{0.3cm} , \hspace{0.3cm}
\left[ \, t^{a} \, , \, t^{b} \, \right]=i \, f^{abc} \, T^{c}
\hspace{0.4cm} \mbox{and} \hspace{0.4cm}
\left[ \, T^{a} \, , \, t^{b} \, \right]=i \, f^{abc} , t^{c} \; ,
\end{equation}
which is just the same Lie algebra, but rewritten in another basis of
generators. Consequently, $\{T^{a}\}$ and $\{t^{a}\}$ are the independents
generators basis for symmetry $SU(N)\times SU(N)$.

Now one needs to obtain the symmetry transformations for the physical fields
$\left( \, G^{\mu} \, , \, C^{\mu} , \right)$, then we consider the transformation
\begin{eqnarray}\label{transfDcovGCchi}
D_{\mu}(G,C)^{\prime}=U \, D_{\mu}(G,C) \, U^{-1} \; ,
\end{eqnarray}
and using the equations above, one case is interested for us in which one
chooses $\omega_{1}^{\;a}=\omega_{2}^{\;a}=\omega^{a}$, it seems to write
the above transformation as solution in terms of transformations for
$\left( \, G^{\mu} \, , \, C^{\mu} \, \right)$
\begin{eqnarray}\label{transfgaugeGCSep}
G_{\mu}^{\;\prime}=U \, G_{\mu} \, U^{-1}+\frac{i}{g_1} \, \left(\partial_{\mu}U\right) \, U^{-1}
\hspace{0.5cm} \mbox{and} \hspace{0.5cm}
C_{\mu}^{\;\prime}=U \, C_{\mu} \, U^{-1}
\; .
\end{eqnarray}
The infinitesimal transformations of the components $\left( \, G_{\mu}^{\;a}\, , \, C_{\mu}^{\;a} \, \right)$ are
\begin{eqnarray}\label{transfgaugeGCSepinf}
G_{\mu}^{\; \prime \; a}=G_{\mu}^{\; a}+f^{abc} \, G_{\mu}^{\; b} \, \omega^{c}-%
\frac{1}{g_{1}} \, \partial_{\mu}\omega^{\;a}
\hspace{0.5cm} \mbox{and} \hspace{0.5cm}
C_{\mu}^{\; \prime \; a}=C_{\mu}^{\; a}+f^{abc} \, C_{\mu}^{\;b} \, \omega^{c} \; .
\end{eqnarray}
The first transformation of (\ref{transfgaugeGCSep}) is just the usual case
of a non-abelian gauge field. The second for $C_{\mu}$ is an unitary and
homogeneous transformation that has interpretation of rotation of the field
$C_{\mu}$ in the isospin space of the composite group $SU(N)\times SU(N)$.
We will analyze the consequences of those transformations dictated by
symmetry $SU(N)\times SU(N)$.

The case above mentioned $\omega _{1}^{\;a}=\omega _{2}^{\;a}=\omega ^{a}$
it is important in order to introduce a mass term for the vector field $C_{\mu }$
into the lagrangian by respecting the invariance principle
dictated by transformations (\ref{transfgaugeGCSep}), while the field $G_{\mu }$
remains massless like in the usual Yang-Mills symmetry. Therefore
these model shows the introduction of a mass term with no necessity to
establish a sector of Higgs scalar field.

Perhaps at this moment it should be more precise to change nomenclature.
Instead of non-abelian $SU(N)\times SU(N)$ to consider a double $SU(N)$
under a common gauge parameter $\omega^{a}$. By double $SU(N)$ one means to
consider fields $\psi_{i}$ and $\phi_{i}$ given by (\ref{quarkcomp},\ref%
{transfglobais12}) as two fundamental representation rotating under the same
group. One step ahead, we should also say that, by identifying $%
\omega_{1}^{a}$ and $\omega_{2}^{a}$, the two quantum numbers of the
different $SU(N)$ factors collapse into a unique quantum number. This then
means that $\psi_{i}$ and $\phi_{i}$ carry the same $SU(N)$-quantum number.
Prior to identification, $\psi_{i}$ and $\phi_{i}$ carrying quantum numbers
of different natures. With $\omega_{1}^{a}=\omega_{2}^{a}=\omega^{\, a}$, the composite
field $\chi$ and its charge is given by a combination of charges of same
nature. In other words : $\psi_{i}$, $\phi_{i}$ and $\chi_{i}$ have all the
same type of color, though their colours have different values.

For constructing the most general lagrangian invariant by those
transformations (\ref{transfgaugeGCSep}) one catalogues all tensors that
transform as
\begin{equation}\label{TransfT}
T \, \longmapsto \, T^{\prime }=U \, T \, U^{-1} \; ,
\end{equation}
like the strength field tensor $F^{\mu \nu }$ in usual Yang-Mills symmetry.
At this way one derives the following fields strength tensors
\begin{equation}\label{TensorsFfC}
i \, g_{1} \, F_{\mu \nu}(G)=\left[ \, D_{\mu }(G) \, , \, D_{\nu }(G)\, \right] \; ,
\hspace{0.3cm}
f_{\mu \nu}(G,C)=\left[ \, D_{\mu }(G) \, , \, C_{\nu} \, \right]
\hspace{0.3cm} \mbox{and} \hspace{0.3cm}
C_{\mu\nu}(C)=g_{3} \, C_{\mu} \, C_{\nu } \; ,
\end{equation}
where we have defined the usual covariant derivative
\begin{equation}\label{DcovG}
D_{\mu }(G)=\partial _{\mu }+i \, g_{1} \, G_{\mu } \; .
\end{equation}
By rewriting those tensors into the basis generators components notation we
obtain
\begin{equation}\label{Fmunu}
F_{\mu \nu }=F_{\mu \nu }^{a} \, T^{a}
\hspace{0.4cm} \mbox{where} \hspace{0.4cm}%
F_{\mu \nu }^{\;a}=\partial _{\mu }G_{\nu }^{a}-\partial _{\nu }G_{\mu
}^{a}-g_{1} \, f^{abc} \, G_{\mu }^{b} \, G_{\nu }^{c} \; .
\end{equation}
The second tensor has the mix between $G^{\mu}$ and $C^{\mu}$
\begin{equation}\label{fmunu}
f_{\mu \nu }=f_{\mu \nu }^{a} \, t^{a} \; ,
\hspace{0.3cm}\mbox{with}\hspace{0.3cm}
f_{\mu \nu }^{a}=\partial _{\mu }C_{\nu }^{a}-g_{1} \, f^{abc} \, G_{\mu }^{b} \, C_{\nu}^{c} \; ,
\end{equation}
in which it is split into the antisymmetric and symmetric parts
\begin{equation}\label{fmunuantisym}
f_{[\mu \nu ]}^{ \, a}=\partial _{\mu }C_{\nu }^{a}-\partial _{\nu }C_{\mu
}^{a}-g_{1} \, f^{abc} \, G_{\mu }^{b} \, C_{\nu }^{c}-g_{1} \, f^{abc} \, C_{\mu }^{b} \,
G_{\nu}^{c} \; ,
\end{equation}
and
\begin{equation}\label{fmunusym}
f_{(\mu \nu )}^{a}=\partial _{\mu }C_{\nu }^{a}+\partial _{\nu }C_{\mu
}^{a}-g_{1} \, f^{abc} \, G_{\mu }^{b} \, C_{\nu }^{c}+g_{1} \, f^{abc} \, C_{\mu }^{b} \, G_{\nu
}^{c} \; ,
\end{equation}
respectively. The symmetrical part of $f^{\mu \nu }$ reveals a longitudinal
propagation for the field $C^{\mu }$ beyond a transversal as a consequence
of transformations (\ref{transfgaugeGCSep}). Indeed, the vector field $%
C^{\mu}$ can be interpreted as a Proca field with a longitudinal
propagation beyond a transversal one. The third tensor is defined only in
terms of $C^{\mu}$ for antisymmetric part

\begin{equation}\label{Cmunuantisym}
C_{[\mu \nu ]}=C_{[\mu \nu ]}^{\;a} \, T^{a}
\; ,
\hspace{0.5cm} \mbox{with} \hspace{0.5cm}
C_{[\mu \nu ]}^{a}=g_{3} \, f^{abc} \, C_{\mu }^{b} \, C_{\nu }^{c} \; ,
\end{equation}
and the symmetric part
\begin{equation}\label{Cmunusym}
C_{(\mu \nu )}=g_{3} \, \left\{ \, C_{\mu } \, , \, C_{\nu } \, \right\}=
g_{3} \, C_{\mu}^{\;a} \, C_{\nu }^{\;b} \, \left\{ \, t^{a} \, , \, t^{b} \,\right\} =g_{3} \, C_{\mu }^{\;a} \, C_{\nu}^{\;b}
\, \left( \, 4 \, \delta ^{ab} \, {\bf 1}-2 \, t_{1}^{a}\otimes t_{2}^{b}-2 \, t_{1}^{b} \otimes t_{2}^{a}+d^{abc} \, T^{c} \, \right) \; ,
\end{equation}
in which $g_{3}$ is the constant coupling associated to self-interactions of
massive non-abelian gauge field.

Now we can define a general tensor $Z_{\mu \nu }$ as linear combination of
the tensors defined in (\ref{TensorsFfC}), and we split in their
antisymmetric and symmetric parts. Therefore we split $Z_{\mu \nu }$ in
antisymmetric and symmetric parts
\begin{equation}\label{Zmunuantisymesym}
Z_{[\mu \nu ]}=F_{\mu \nu }+a \, f_{[\mu \nu ]}+b \, C_{[\mu \nu ]}
\hspace{0.5cm} \mbox{and} \hspace{0.5cm}
Z_{(\mu\nu)}=c \, f_{(\mu\nu)}+d \, z_{(\mu\nu)}
+\eta_{\mu\nu} \, \left( \, e \, f_{(\alpha}^{\;\;\; \alpha)}
+f \, z_{(\alpha}^{\;\;\; \alpha)} \, \right) \; ,
\end{equation}
in which $\left(a,b,c,d,e,f\right)$ are real parameters, thus
\begin{equation}\label{ZmunuTZmunut}
Z_{[\mu \nu ]}=Z_{[\mu \nu ]}^{(T)a} \, T^{a}+Z_{[\mu \nu ]}^{(t)a} \, t^{a} \; ,
\end{equation}
where
\begin{equation}\label{ZmunuFfC}
Z_{[\mu \nu ]}^{(T)a}=F_{\mu \nu }^{a}+b \, C_{[\mu \nu ]}^{a}
\hspace{0.7cm} \mbox{and} \hspace{0.7cm}
Z_{[\mu \nu ]}^{(t)a}=a \, f_{[\mu \nu ]}^{a} \; .
\end{equation}
Similarly,
\begin{equation}\label{ZmunuAntisym}
Z_{(\mu \nu )}=Z_{(\mu \nu )}^{(T)a} \, T^{a}+Z_{(\mu \nu )}^{(t)a} \, t^{a}+Z_{(\mu\nu )}^{(\Lambda )ab} \, \Lambda^{ab} \; ,
\end{equation}
where
\begin{eqnarray}\label{ZmunuSimT}
Z_{(\mu \nu )}^{(T)a}=d \, g_{3} \, d^{abc} \, C_{\mu }^{b} \, C_{\nu }^{c}
+f \, g_{3} \, d^{abc} \, \eta_{\mu\nu} \, C_{\alpha}^{b} \, C^{\alpha c}
\; , \hspace{0.3cm}
Z_{(\mu \nu )}^{(t)a}=c \, f_{(\mu \nu )}^{a}+e \, \eta_{\mu\nu} \, f_{(\alpha}^{\;\;\; \alpha)a}
\; , \hspace{0.3cm}
\nonumber \\
Z_{(\mu \nu)}^{(\Lambda )\;ab}=d \, g_{3} \, C_{\mu }^{a} \, C_{\nu }^{b}+f \, g_{3} \, \eta_{\mu\nu} \, C_{\alpha }^{a} \, C^{\alpha b}
\hspace{0.3cm} \mbox{and} \hspace{0.3cm}
\Lambda ^{ab} =4 \, \delta^{ab} \, {\bf 1} -2 \, t_{1}^{a}\otimes
t_{2}^{b}-2t_{1}^{b}\otimes t_{2}^{a} \; .
\end{eqnarray}
The real parameters $(a,b,c,d,e,f)$ have introduced for a better control of all
terms that contribute into the lagrangian, it will be important when one
analyze the aspect of unitarity directly from the vector propagators. Here
it was been convenient to define another base $\{\Lambda ^{ab}\}$, beyond
$\{T^{a}\}$ and $\{t^{a}\}$, in which it satisfy to commutation relations
\begin{eqnarray}\label{TtLambda}
\left[ \, T^{a} \, , \, \Lambda^{bc} \, \right] \!\!\! &=& \!\!\! -2i \, \left[ \, f^{abd} \, \left( \, t_{1}^{d}\otimes
t_{2}^{c}+t_{1}^{c}\otimes t_{2}^{d} \, \right)+f^{acd} \, \left( t_{1}^{b}\otimes
t_{2}^{d}+t_{1}^{d}\otimes t_{2}^{b} \, \right)\right] \; ,  \notag
\label{ComutacoesTtLambda} \\
\left[ \, t^{a} \, , \, \Lambda ^{bc} \, \right] \!\!\!&=&\!\!\! -2i \, \left[ \, f^{abd} \, \left( \, t_{1}^{d}\otimes
t_{2}^{c}-t_{1}^{c} \otimes t_{2}^{d} \, \right)+f^{acd} \, \left( \, t_{1}^{b}\otimes
t_{2}^{d}-t_{1}^{d}\otimes t_{2}^{b} \, \right) \,\right] \; .
\end{eqnarray}

Now we are enable to write the most complete lagrangian invariant by
symmetry transformations (\ref{transfgaugeGCSep}) as
\begin{equation}\label{lagrangianoZmunu}
\mathcal{L}=-\frac{1}{4} \, \mbox{tr}\left( \, Z_{\mu\nu} \, Z^{\mu\nu} \, \right)
+\frac{1}{2} \, m^2 \, \mbox{tr}\left( \, C_{\mu} \, C^{\mu} \, \right)-\frac{1}{2\xi} \, \mbox{tr}\left( \, \partial_{\mu}G_{\mu} \, \right)^{2}
-\frac{1}{4} \, \eta \, \, \mbox{tr}\left( \, \tilde{Z}_{\mu\nu} \, Z^{\mu\nu} \, \right) \; ,
\end{equation}
where we have taken into account the semi-topological term
$\tilde{Z}_{\mu\nu}=\frac{1}{2} \; \varepsilon_{\mu\nu\alpha\beta}Z^{\alpha\beta}$,
that give us a non-trivial contribution and invariant by (\ref{transfgaugeGCSep}
), and the lasts terms are contributions that is not presents in the earlier
definitions of the tensors (\ref{TensorsFfC})-(\ref{ZmunuSimT}),
with $\lambda_{1} \in {\mathbb R}$. Clearly, the semi-topological term brings out a CP violation in the model, but on general aspects we look for complete symmetry with the presence of all terms. The parameter $\eta$ of (\ref{lagrangianoZmunu})
sets a parity violating regime. The mass term for gauge field $C^{\mu}$ has been introduced due to transformations (\ref{transfgaugeGCSep}),
and one has chosen the gauge fixing term
\begin{eqnarray}\label{gaugefixing}
\mathcal{L}_{gf}=-\frac{1}{2\xi} \, \left( \, \partial_{\mu}G^{\mu a} \, \right)^{2} \; ,
\end{eqnarray}
convenient to quantize the model hereafter, where $\xi \in {\mathbb R}$. Using the
traces relations
\begin{eqnarray}\label{tracesTt}
\mbox{tr}(T^{a}T^{b})=\mbox{tr}(t^{a}t^{b})=N \, \delta^{ab}
\hspace{0.2cm} , \hspace{0.2cm}
\mbox{tr}(T^{a}t^{b})=\mbox{tr}(T^{a}\Lambda^{bc})=\mbox{tr}
(t^{a}\Lambda^{bc})=0
\nonumber \\
\mbox{and} \hspace{0.5cm} \mbox{tr}(\Lambda^{ab}\Lambda^{cd})=4 \, \delta^{ab} \,
\delta^{cd} +2 \, \delta^{ac} \, \delta^{bd}+2 \, \delta^{ad} \, \delta^{bc} \; , \hspace{2cm}
\end{eqnarray}
we can obtain all the free and interaction terms from the lagrangian (\ref{lagrangianoZmunu}).
The lagrangian free part of the vector fields $\left( \, G^{\mu} \, , \, C^{\mu} \, \right)$ is given by
\begin{eqnarray}\label{LfreeGC}
\mathcal{L}_{gauge-G-0} =-\frac{1}{4} \, \left( \, \partial _{\mu }G_{\nu }^{\;a}-\partial
_{\nu }G_{\mu }^{\;a} \, \right)^{2}-\frac{1}{2\xi }\, \left( \, \partial _{\mu }G^{\mu
a} \, \right)^{2}
\hspace{0.5cm} \mbox{and} \hspace{4cm}
\nonumber \\
\mathcal{L}_{gauge-C-0} =-\frac{a^{2}}{4} \, \left( \, \partial _{\mu }C_{\nu }^{\;a}-\partial
_{\nu }C_{\mu }^{\;a} \, \right)^{2}-\frac{c^{2}}{4} \, \left( \, \partial _{\mu }C_{\nu
}^{\;a}+\partial _{\nu }C_{\mu }^{\;a} \, \right)^{2}-2 \, e \, \left(c+2 \, e\right) \left( \, \partial_{\mu}C^{\mu a} \, \right)^{2}
+\frac{1}{2} \, m^{2} \, C_{\mu}^{\;a} \, C^{\mu a} \; ,
\end{eqnarray}
respectively. The Faddeev-Popov ghost lagrangian can be added by the usual methods using
the infinitesimal transformations (\ref{transfgaugeGCSepinf}) and the gauge
fixing term (\ref{gaugefixing}), then one gets
\begin{equation}\label{Lfaddeev}
\mathcal{L}_{FP}=\bar{\eta}^{a} \, \left( \, \delta ^{ab}\Box \, \right)\eta
^{b}+g_{1} \, f^{abc} \, \partial _{\mu }\bar{\eta}^{a} \, G^{\mu c} \, \eta ^{b} \; ,
\end{equation}
where $\left( \, \bar{\eta} \, , \, \eta \, \right)$ are Faddeev-Popov fields. From the lagrangian free part we calculate the $\left( \, G_{\mu}^{a} \, , \, C_{\mu }^{a} \, \right)$ vector fields propagators by writing it into the form
\begin{eqnarray}\label{L0GCoperators}
\mathcal{L}_{gauge-0} \!\!\!&=&\!\!\! \frac{1}{2} \, G^{\mu a} \, \left( \, \Box \, \theta_{\mu\nu}+\xi^{-1} \, \Box \, \omega_{\mu\nu} \, \right) \, G^{\nu a}+
\nonumber \\
&&
\hspace{-0.5cm}
+\frac{1}{2} \, C^{\mu a} \, \left[ \, \left(\, \left(a^{2}+c^{2}\right) \, \Box+m^{2} \right) \, \theta_{\mu\nu}+\left( \, \left(2 \, c^{2}+2 \, e \, (c+2e) \, \right) \, \Box+m^{2}\right) \, \omega_{\mu\nu} \, \right] \, C^{\nu a} \; ,
\end{eqnarray}
in terms of projection operators
\begin{equation}\label{thetaomegamunu}
\theta_{\mu \nu }=\eta_{\mu \nu }-\frac{\partial _{\mu } \, \partial _{\nu }}{\Box}
\hspace{0.5cm}\mbox{and}\hspace{0.5cm}
\omega_{\mu \nu }=\frac{\partial_{\mu } \, \partial _{\nu }}{\Box } \; .
\end{equation}
We invert the operators from (\ref{L0GCoperators}) between fields to obtain the
propagators in the momentum space. In the case of massless gauge fields is
exactly like the propagator in Yang-Mills. For massive vector fields one
gets
\begin{equation}\label{PropagatorC}
\langle \, C_{\mu }^{\;a} \, C_{\nu }^{\;b} \, \rangle =-i \, \delta ^{ab} \, \left[ \frac{1}{
\left(a^{2}+c^{2}\right) \, k^{2}-m^{2}}\left( \, \eta_{\mu \nu }-\frac{k_{\mu } \, k_{\nu }}{k^{2}} \, \right) +\frac{1}{\left( \, 2c^{2}+2e(c+2e) \, \right) \, k^{2}-m^{2}} \, \frac{k_{\mu} \, k_{\nu }}{k^{2}} \, \right] \; ,
\end{equation}
in which one has observed two masses
\begin{equation}\label{massasmu}
\mu_{1}^{\, 2}=\frac{m^{2}}{a^{2}+c^{2}}
\hspace{0.5cm} \mbox{and} \hspace{0.5cm}
\mu_{2}^{\, 2}=\frac{m^{2}}{2 \, c^{2}+2 \, e \, (c+2 \, e)} \; ,
\end{equation}
for transversal part of spin-$1$, and another mass for longitudinal part of
spin-$0$. The expressions of all propagators are showed in the figure
(\ref{propagators}).
%
%
%
%
\begin{figure}[!h]
\begin{center}
\newpsobject{showgrid}{psgrid}{subgriddiv=1,griddots=10,gridlabels=6pt}
\begin{pspicture}(-2,-1)(8,2)
%
\pscoil[coilarm=0,coilaspect=0,coilwidth=0.2,coilheight=1.0,linecolor=black](5,1)(7,1)
%
%
\put(-2.5,1){$\langle \, G_{\mu}^{\;a} \, G_{\nu}^{\;b} \, \rangle=-\frac{i \, \delta^{ab}}{k^2}
\left[ \, \eta_{\mu\nu}+\left( \xi-1 \right) \, \frac{k_{\mu} \, k_{\nu}}{k^2} \, \right]$}
%
%
\put(4.9,1.25){$\mu$}
\put(6.9,1.25){$\nu$}
\put(5,0.5){$a$}
\put(7,0.5){$b$}
\put(6,0.4){$k$}
%
%
%
\end{pspicture}
\end{center}
\end{figure}
%
%
%
\begin{figure}[!h]
\begin{center}
\newpsobject{showgrid}{psgrid}{subgriddiv=1,griddots=10,gridlabels=6pt}
\begin{pspicture}(-2,0)(8,0.7)
%
\pscoil[coilarm=0,coilwidth=0.2,coilheight=1.0,linecolor=black](8,2)(10,2)
%
%
\put(-5,2){$\langle \, C_{\mu}^{\;a} \, C_{\nu}^{\;b} \, \rangle=-\frac{i\delta^{ab}}{\left(a^2+c^2\right) \, k^2-m^2}
\left[ \, \eta_{\mu\nu}+\left(a^2-c^2-2 \, e \, \left( c+2 \, e \right)\right) \, \frac{k_{\mu} \, k_{\nu}}{\left( \, 2 \, c^2+2 \, e(c+2e) \, \right) \, k^2-m^2}\right]$}
%
%
\put(7.9,2.25){$\mu$}
\put(9.85,2.25){$\nu$}
\put(8,1.5){$a$}
\put(10,1.5){$b$}
\put(9,1.3){$k$}
%
%
%
\end{pspicture}
\end{center}
\end{figure}
%
%
%
\begin{figure}[!h]
\begin{center}
\newpsobject{showgrid}{psgrid}{subgriddiv=1,griddots=10,gridlabels=6pt}
\begin{pspicture}(0,0)(8,0.5)
%
\psline[linestyle=dashed,linewidth=0.1,linecolor=black](3.5,2)(5.5,2)
%
%
\put(-0.5,2){$\langle \, \bar{\eta}^{a} \, \eta^{b} \, \rangle=-i \, \frac{\delta^{ab}}{p^2}$}
%
%
\put(3.4,2.2){$a$}
\put(5.4,2.2){$b$}
\put(4.45,1.5){$p$}
%
%
%
\end{pspicture}
\vspace{-1.5cm}
\end{center}
\caption{Feynman propagators of the vector fields and Ghosts.}
\label{propagators}
\end{figure}
%
%
%


In order to show an initial consistency of the model one should to observe
the behavior of those propagators when $k \longrightarrow \infty$, it goes
to zero in ultraviolet regime.
Differently from usual Proca's case, the massive field $C_{\mu}$ has a
health behavior. Considering the unitarity one guarantees the positivity of
the propagators residue if one imposes the following inequalities :
$a^{2}+c^{2} > 1$ and $2 \, c^2+2 \, c \, e+4 \, e^{2}<1$, on those conditions the model is
unitary at the tree level. Since $(a,c)$ are real parameters, by imposing
the above conditions there is no any possibility for emergence of tachyons
propagation here.

It is also interesting to see the independency of the massive field
propagator in relation to any parameters if we had introduced the most
general gauge fixing
\begin{equation}\label{LgaugefixGC}
\mathcal{L}_{gf}=-\frac{1}{2\xi} \, \left( \, \partial_{\mu}G^{\mu a}+\sigma \, \partial_{\mu}C^{\mu
a} \, \right)^{2} \; ,
\end{equation}
with $\sigma \in {\mathbb R}$.


The vector fields and Faddeev-Popov interaction terms are given through the
lagrangian terms
\begin{eqnarray}\label{LInt3}
\mathcal{L}_{gauge-int}^{\;(3)} \!\!\!&=&\!\!\! g_{1} \, \partial_{\mu}G_{\nu}^{\;a} \, \left[ \, G^{\mu} \, , \, G^{\nu} \, \right]^{a}
+ a^2 \, g_{1} \, \partial_{\mu}C_{\nu}^{\;a} \, \left(\left[ \, G^{\mu} \, , \, C^{\nu} \, \right]^{a}
+\left[ \, C^{\mu} \, , \, G^{\nu} \, \right]^{a}\right)
\nonumber \\
&&
\hspace{-0.5cm}
+ \, b \, g_{3} \, \partial_{\mu}G_{\nu}^{\; a} \, \left[ \, C^{\mu} \, , \, C^{\nu} \, \right]^{a}
+c^2 \, g_{1} \, \partial_{\mu}C_{\nu}^{\; a}
\left(\left[ \, G^{\mu} \, , \, C^{\nu} \, \right]^{a}-\left[ \, C^{\mu} \, , \, G^{\nu} \, \right]^{a}\right)
\nonumber \\
&&
\hspace{-0.5cm}
+ \, 4 \, e \, \left(c+e\right) \, g_{1} \, \partial_{\mu}C^{\mu a} \, \left[ \, G_{\nu} \, , \, C^{\nu} \, \right]^{a} \; ,
\end{eqnarray}
and
\begin{eqnarray}\label{LKInt4}
\mathcal{L}_{gauge-int}^{\; (4)} \!\!\!&=&\!\!\! -\frac{1}{4} \, g_{1}^{2} \, \left[ \, G_{\mu} \, , \, G_{\nu} \, \right]^{a}
\left[ \, G^{\mu} \, , \, G^{\nu} \, \right]^{a}-\frac{a^2}{2} \, g_{1}^{2} \, \left[ \, G_{\mu} \, , \, C_{\nu} \, \right]^{a}
\left(\left[ \, G^{\mu} \, , \, C^{\nu} \, \right]^{a}+\left[ \, C^{\mu} \, , \, G^{\nu} \, \right]^{a}\right)
\nonumber \\
&&
\hspace{-0.5cm}
-\frac{b}{2} \, g_{1} \, g_{3} \, \left[ \, G_{\mu} \, , \, G_{\nu} \, \right]^{a} \, \left[ \, C^{\mu} \, , \, C^{\nu} \, \right]^{a}
-\frac{b^2}{4} \, g_{3}^{2} \, \left[ \, C_{\mu} \, , \, C_{\nu} \, \right]^{a} \, \left[ \, C^{\mu} \, , \, C^{\nu} \, \right]^{a}
\nonumber \\
&&
\hspace{-0.5cm}
-\frac{c^2}{2} \, g_{1}^{2} \, \left[ \, G_{\mu} \, , \, C_{\nu} \, \right]^{a}
\left( \left[ \, G^{\mu} \, , \, C^{\nu} \, \right]^{a}-\left[ \, C^{\mu} \, , \, G^{\nu} \, \right]^{a}\right)
-4 \, e^{2} \, g_{1}^{2} \, \left[ \, G_{\mu} \, , \, C^{\mu} \, \right]^{a} \left[ \, G_{\nu} \, , \, C^{\nu} \, \right]^{a}
\nonumber \\
&&
\hspace{-0.5cm}
-\frac{1}{N}\left[\frac{3d^2}{2}+2f\left(d+2f\right)\right] \, g_{3}^{2} \, C_{\mu}^{a} \, C_{\nu}^{a} \, C^{\mu b} \, C^{\nu b}
-\frac{1}{N}\left[\frac{d^2}{2}+2f\left(d+2f\right)\right] \, g_{3}^{2} \, C_{\mu}^{a} \, C_{\nu}^{b} \, C^{\mu a} \, C^{\nu b}
\nonumber \\
&&
\hspace{-0.5cm}
-\frac{d^2}{4} \, g_{3}^{2} \, d^{abc} \, d^{ade} \, C_{\mu}^{\;b} \, C_{\nu}^{\; c} \, C^{\mu d} \, C^{\nu e}
-\left(\frac{d^2}{4}+f^{2}\right)g_{3}^{\, 2} \, d^{abc} \, d^{ade} \, C_{\mu}^{\;b} \, C^{\mu c} \, C_{\nu}^{\;d} \, C^{\nu e} \; .
\end{eqnarray}
for the three-line and four-line vertex, respectively.
The Feynman rules for vertex are obtained in the momentum space :
%
%
%
\begin{figure}[h]
\begin{center}
\newpsobject{showgrid}{psgrid}{subgriddiv=1,griddots=10,gridlabels=6pt}
\begin{pspicture}(0,0)(13,2.6)
%
\pscoil[coilarm=0,coilaspect=0,coilwidth=0.2,coilheight=1.0,linecolor=black](0,1)(0,2.5)
\pscoil[coilarm=0,coilaspect=0,coilwidth=0.2,coilheight=1.0,linecolor=black](0,1)(1,0)
\pscoil[coilarm=0,coilaspect=0,coilwidth=0.2,coilheight=1.0,linecolor=black](0,1)(-1,0)
%
%
%
\put(0.3,2.5){$a,\mu$}
\put(-1.8,0){$b,\nu$}
\put(1.3,0){$c,\rho$}
\put(0.3,1.7){$k_{1}$}
\put(-1,0.7){$k_{2}$}
\put(0.7,0.7){$k_{3}$}
%
%
\put(1.5,1){$i \, V^{(3)abc}_{(G)\mu\nu\rho}(k_{1},k_{2},k_{3})=
g_{1} \, f^{abc} \, \left[ \, \eta_{\mu\nu}(k_{1}-k_{2})_{\rho}+\eta_{\nu\rho}(k_{2}-k_{3})_{\mu}
+\eta_{\mu\rho}(k_{3}-k_{1})_{\nu} \, \right]$}
\end{pspicture}
\end{center}
\end{figure}
%
%
\begin{figure}[h]
\begin{center}
\newpsobject{showgrid}{psgrid}{subgriddiv=1,griddots=10,gridlabels=6pt}
\begin{pspicture}(0,0)(13,2)
%
\pscoil[coilarm=0,coilaspect=0,coilwidth=0.2,coilheight=1.0,linecolor=black](0,1)(1,2)
\pscoil[coilarm=0,coilaspect=0,coilwidth=0.2,coilheight=1.0,linecolor=black](0,1)(1,0)
\pscoil[coilarm=0,coilaspect=0,coilwidth=0.2,coilheight=1.0,linecolor=black](0,1)(-1,0)
\pscoil[coilarm=0,coilaspect=0,coilwidth=0.2,coilheight=1.0,linecolor=black](0,1)(-1,2)
%
%
\put(-1.8,2){$a,\mu$}
\put(1.3,2){$b,\nu$}
\put(-1.8,0){$c,\rho$}
\put(1.3,0){$d,\sigma$}
\put(-0.5,1.8){$k_{1}$}
\put(-1,0.7){$k_{3}$}
\put(0.25,1.8){$k_{2}$}
\put(0.6,0.7){$k_{4}$}
%
%
\put(2.5,1){$i \, V^{(4)abcd}_{(G)\mu\nu\rho\sigma}=-i \, g_{1}^{2} \,
\; \left[ \, f^{eab} \, f^{ecd} \, \left(\eta_{\mu\rho} \, \eta_{\nu\sigma}-\eta_{\mu\sigma} \, \eta_{\nu\rho}\right)\right.$}
\put(2.5,0){$\left.+f^{eac} \, f^{ebd} \, \left( \eta_{\mu\nu} \, \eta_{\rho\sigma}-\eta_{\mu\sigma} \, \eta_{\rho\nu} \right)
+f^{ead} \, f^{ebc} \, \left( \, \eta_{\mu\nu} \, \eta_{\rho\sigma}-\eta_{\mu\rho} \, \eta_{\nu\sigma} \right) \, \right]$}
\end{pspicture}
\vspace{0.3cm}
\end{center}
\caption{Three and four lines vertices for massless gauge fields. It is just
usual Yang-Mills vertices.}
\label{VertexQCD}
\end{figure}
%
%
%
%
\begin{figure}[h]
\begin{center}
\newpsobject{showgrid}{psgrid}{subgriddiv=1,griddots=10,gridlabels=6pt}
\begin{pspicture}(0,-1)(13,1.5)
\psset{arrowsize=0.2 2}
%
\pscoil[coilarm=0,coilwidth=0.2,coilheight=1.0,linecolor=black](0,0)(0,1.5)
\pscoil[coilarm=0,coilwidth=0.2,coilheight=1.0,linecolor=black](0,0)(1,-1)
\pscoil[coilarm=0,coilaspect=0,coilwidth=0.2,coilheight=1.0,linecolor=black](0,0)(-1,-1)
\put(0.3,1.5){$a,\mu$}
\put(-1.8,-1){$b,\nu$}
\put(1.3,-1){$c,\rho$}
\put(0.3,0.8){$k_{1}$}
\put(-1,-0.3){$k_{2}$}
\put(0.7,-0.3){$k_{3}$}
\put(2.5,0){$i \, V^{(3)abc}_{(GC)\mu\nu\rho}(k_{1},k_{2},k_{3})=ib \, g_{3} \, f^{abc}
\left( \eta_{\nu\rho} \, k_{2\mu}-\eta_{\mu\nu} \, k_{2\rho}\right)$}
\put(2.5,-1){$+(a^2+c^2) \, g_{1} \, f^{abc} \, \eta_{\mu\rho}(k_{1}-k_{3})_{\nu}
+(a^2-c^2)g_{1}f^{abc}\left(\eta_{\mu\rho} \, k_{1\nu}-\eta_{\nu\rho} \, k_{2\mu}\right)$}
\put(2.5,-2){$+4 \, e(c+e) \, g_{1} \, f^{abc} \, \left( \eta_{\nu\rho} \, k_{1\mu}-\eta_{\mu\nu} \, k_{3\rho}\right)$}
\end{pspicture}
\end{center}
\end{figure}
%
%
%
\begin{figure}[h]
\begin{center}
\newpsobject{showgrid}{psgrid}{subgriddiv=1,griddots=10,gridlabels=6pt}
\begin{pspicture}(4,-1)(9,2)
%
%
\pscoil[coilarm=0,coilaspect=0,coilwidth=0.2,coilheight=1.0,linecolor=black](0,1)(1,2)
\pscoil[coilarm=0,coilwidth=0.2,coilheight=1.0,linecolor=black](0,1)(1,0)
\pscoil[coilarm=0,coilwidth=0.2,coilheight=1.0,linecolor=black](0,1)(-1,0)
\pscoil[coilarm=0,coilaspect=0,coilwidth=0.2,coilheight=1.0,linecolor=black](0,1)(-1,2)
\put(-1.8,2){$a,\mu$}
\put(1.3,2){$b,\nu$}
\put(-1.8,0){$c,\rho$}
\put(1.3,0){$d,\sigma$}
\put(-0.5,1.8){$k_{1}$}
\put(-0.5,0){$k_{3}$}
\put(0.2,1.8){$k_{2}$}
\put(0.3,0){$k_{4}$}
\put(2.5,1){$i \, V^{(4)abcd}_{(GC)\mu\nu\rho\sigma} = i \, b \, g_{3} \, g_{1} \, f^{eab} \, f^{ecd} (\eta_{\mu\rho} \, \eta_{\nu\sigma}-\eta_{\mu\sigma} \, \eta_{\nu\rho})$}
\put(2.5,0){$-(a^2+c^2) \, i \, g_{1}^{2} \; \eta_{\mu\nu} \, \eta_{\rho\sigma}(f^{eac} \, f^{ebd}+f^{ead} \, f^{ebc})$}
\put(2.5,-1){$ +(a^2-c^2) \, i \, g_{1}^{2}
\;(f^{eac} \, f^{ebd} \, \eta_{\mu\sigma} \, \eta_{\nu\rho}+f^{ead} \, f^{ebc} \, \eta_{\mu\rho} \, \eta_{\nu\sigma})$}
\put(2.5,-2){$ -i \, 8 \, e^{2} \, g_{1}^{2}
\;\left(f^{eab} \, f^{ecd} \, \eta_{\mu\nu} \, \eta_{\rho\sigma}-f^{ead} \, f^{ebc} \, \eta_{\mu\sigma} \, \eta_{\nu\rho} \, \right)$}
\end{pspicture}
\end{center}
\end{figure}
%
%
%
%
\begin{figure}[h]
\begin{center}
\newpsobject{showgrid}{psgrid}{subgriddiv=1,griddots=10,gridlabels=6pt}
\begin{pspicture}(1,0)(11,3.5)
%
\pscoil[coilarm=0,coilwidth=0.2,coilheight=1.0,linecolor=black](0,1)(1,2)
\pscoil[coilarm=0,coilwidth=0.2,coilheight=1.0,linecolor=black](0,1)(1,0)
\pscoil[coilarm=0,coilwidth=0.2,coilheight=1.0,linecolor=black](0,1)(-1,0)
\pscoil[coilarm=0,coilwidth=0.2,coilheight=1.0,linecolor=black](0,1)(-1,2)
%
%
\put(-1.8,2){$a,\mu$}
\put(1.3,2){$b,\nu$}
\put(-1.8,0){$c,\rho$}
\put(1.3,0){$d,\sigma$}
\put(-0.5,1.8){$k_{1}$}
\put(-1,0.7){$k_{3}$}
\put(0.25,1.8){$k_{2}$}
\put(0.6,0.7){$k_{4}$}
%
%
\put(2.5,1){$i \, V^{(4)abcd}_{(C)\mu\nu\rho\sigma}=-i \, b^2 \, g_{3}^{2}\;
\left[ \, f^{eab} \, f^{ecd} \, \left( \, \eta^{\mu\rho} \, \eta^{\nu\sigma}-\eta^{\mu\sigma} \, \eta^{\nu\rho}\right)\right.$}
\put(4,0){$+f^{eac} \, f^{ebd} \, \left( \eta^{\mu\nu} \, \eta^{\rho\sigma}-\eta^{\mu\sigma} \, \eta^{\rho\nu} \right)$}
\put(-2.5,-1){$\left.+f^{ead} \, f^{ebc}\left( \eta^{\mu\nu} \, \eta^{\rho\sigma}-\eta^{\mu\rho} \, \eta^{\nu\sigma}\right) \right]
-\frac{6d^2+8f(d+f)}{N} \, i \, g_{3}^{2}
\;\left[ \; \delta^{ab} \, \delta^{cd}(\eta^{\mu\rho} \, \eta^{\nu\sigma}+\eta^{\mu\sigma} \, \eta^{\nu\rho})+\delta^{ac}\delta^{bd}
(\eta^{\mu\nu} \, \eta^{\rho\sigma}+\eta^{\mu\sigma} \, \eta^{\nu\rho})\right.$}
\put(-2.5,-2){$+\delta^{ad}\delta^{bc}\left(\eta^{\mu\nu} \, \eta^{\rho\sigma}+\eta^{\mu\rho} \, \eta^{\nu\sigma}\right)]
-\frac{4d^{2}+16f(d+f)}{N} \, i \, g_{3}^{\;2}\;\left( \delta^{ab} \, \delta^{cd} \, \eta^{\mu\nu} \, \eta^{\rho\sigma}+\delta^{ac} \, \delta^{bd}
\eta^{\mu\rho} \, \eta^{\nu\sigma} +\delta^{ad} \, \delta^{bc} \, \eta^{\mu\sigma} \, \eta^{\nu\rho}\right)$}
\put(-2.5,-3){$-d^2 \, i \, g_{3}^{2} \, \left[ \, d^{eab} \, d^{ecd} \, \left( \, \eta^{\mu\rho} \, \eta^{\nu\sigma}+\eta^{\mu\sigma} \, \eta^{\nu\rho}\right)
+d^{eac} \, d^{ebd}\left( \eta^{\mu\nu} \, \eta^{\rho\sigma}+\eta^{\mu\sigma} \, \eta^{\rho\nu}\right)
+d^{ead} \, d^{ebc}\left( \eta^{\mu\nu} \, \eta^{\rho\sigma}+\eta^{\mu\rho} \, \eta^{\nu\sigma} \right) \right]$}
\put(-2.5,-4){$-(d^2+4 \, f^{2}) i \, g_{3}^{2}[d^{eab} \, d^{ecd} \, \left( \eta^{\mu\rho} \, \eta^{\nu\sigma}+\eta^{\mu\sigma} \, \eta^{\nu\rho}\right)
+d^{ead} \, d^{ebc}\left( \eta^{\mu\nu} \, \eta^{\rho\sigma}+\eta^{\mu\sigma} \, \eta^{\rho\nu}\right)
+d^{eac} \, d^{ebd}\left( \eta^{\mu\nu} \, \eta^{\rho\sigma}+\eta^{\mu\rho} \, \eta^{\nu\sigma}\right)]$}
\end{pspicture}
\vspace{3.8cm}
\end{center}
\caption{New vertices of the $SU(N)\times SU(N)$ symmetry mixing massless
and massive vector fields.}
\label{NewVertex}
\end{figure}
%
%
%
%
\begin{figure}[h]
\begin{center}
\newpsobject{showgrid}{psgrid}{subgriddiv=1,griddots=10,gridlabels=6pt}
\begin{pspicture}(0,0)(10,1.5)
\psset{arrowsize=0.2 2}
%
\pscoil[coilarm=0,coilwidth=0.2,coilheight=1.0,linecolor=black](0,0)(0,1.5)
\pscoil[coilarm=0,coilwidth=0.2,coilheight=1.0,linecolor=black](0,0)(1,-1)
\pscoil[coilarm=0,coilaspect=0,coilwidth=0.2,coilheight=1.0,linecolor=black](0,0)(-1,-1)
\put(0.3,1.5){$a,\mu$}
\put(-1.8,-1){$b,\nu$}
\put(1.3,-1){$c,\rho$}
\put(0.3,0.8){$k_{1}$}
\put(-0.9,-0.3){$k_{2}$}
\put(0.7,-0.3){$k_{3}$}
\put(2,0){$i \, \widetilde{V}^{(3)abc}_{(GC)\mu\nu\rho}(k_{1},k_{3})
=a^2 \, g_{1} \, f^{abc} \, \varepsilon_{\alpha\mu\nu\rho} \, \left(k_{1}+k_{3}\right)^{\alpha}$}
\end{pspicture}
\end{center}
\end{figure}
%
%
%
%
\begin{figure}[h]
\begin{center}
\newpsobject{showgrid}{psgrid}{subgriddiv=1,griddots=10,gridlabels=6pt}
\begin{pspicture}(0,0)(10,3.3)
%
\pscoil[coilarm=0,coilaspect=0,coilwidth=0.2,coilheight=1.0,linecolor=black](0,1)(1,2)
\pscoil[coilarm=0,coilwidth=0.2,coilheight=1.0,linecolor=black](0,1)(1,0)
\pscoil[coilarm=0,coilwidth=0.2,coilheight=1.0,linecolor=black](0,1)(-1,0)
\pscoil[coilarm=0,coilaspect=0,coilwidth=0.2,coilheight=1.0,linecolor=black](0,1)(-1,2)
\put(-1.8,2){$a,\mu$}
\put(1.3,2){$b,\nu$}
\put(-1.8,0){$c,\rho$}
\put(1.3,0){$d,\sigma$}
\put(-0.5,1.8){$k_{1}$}
\put(-0.5,0){$k_{3}$}
\put(0.2,1.8){$k_{2}$}
\put(0.3,0){$k_{4}$}
\put(2,1){$i \, \widetilde{V}^{(4)abcd}_{(GC)\mu\nu\rho\sigma}=
-a^2 \, i \, g_{1}^{2} \, \varepsilon_{\mu\nu\rho\sigma}\left( f^{ead} \, f^{ebc}-f^{eac} \, f^{ebd} \right)$}
\end{pspicture}
\end{center}
\caption{Contributions of the semitopological terms.}
\label{SemiTopologicalcontribution}
\end{figure}
%
%
%
%
\begin{figure}[h]
\begin{center}
\newpsobject{showgrid}{psgrid}{subgriddiv=1,griddots=10,gridlabels=6pt}
\begin{pspicture}(0,0)(4,2)
\psset{arrowsize=0.2 2}
%
%
%
%
\pscoil[coilarm=0,coilaspect=0,coilwidth=0.2,coilheight=1.0,linecolor=black](0,0)(0,1.5)
\psline[linestyle=dashed,linewidth=0.5mm]{->}(1,-1)(0.4,-0.4)
\psline[linestyle=dashed,linewidth=0.5mm](0.5,-0.5)(0,0)
\psline[linestyle=dashed,linewidth=0.5mm]{->}(0,0)(-0.7,-0.7)
\psline[linestyle=dashed,linewidth=0.5mm](-0.5,-0.5)(-1,-1)
\put(0.3,1.5){$\mu,c$}
\put(-1.3,-1){$a$}
\put(1.1,-1){$b$}
\put(0.3,0.8){$k$}
\put(-0.8,-0.3){$p^{\prime}$}
\put(0.6,-0.3){$p$}
\put(1.3,0){$V_{\mu}^{(3)\;abc}(p)=g_{1} \, f^{abc} \, p_{\mu}$}
\end{pspicture}
\vspace{1cm}
\end{center}
\caption{The Faddeev-Popov vertex.}
\label{VertexFP}
\end{figure}

\newpage

\subsection{Quarks sector for symmetry $SU_{c}(3) \times SU_{c}(3)$}

In the last subsection we have seen the construction of a composite gauge
symmetry $SU(N)\times SU(N)$ and one has concentred just in the sector of
vector fields $\left( \, G^{\mu } \, , \, C^{\mu} \, \right)$. Here one shall present the fermions
sector to complete the lagrangian (\ref{lagrangianoZmunu}) by adding the
fermionic fields. Our construction initial was based on a fermionic field
$\chi $, that is a composition of a fermion $\psi$ and a scalar $\phi$, but
we will be interested in establishing a dynamic for field $\chi$. Therefore
the lagrangian for fermions sector of $SU(N)\times SU(N)$ is coupling those
fermions represented by $\chi $ to covariant derivative (\ref{DcovGC})
\begin{equation}\label{Lfermions}
\mathcal{L}_{fermions}=\bar{\chi}(x) \, \left[ \, i \, \gamma ^{\mu } \, D_{\mu }(G,C)-m_{\chi} \, \mathbf{1} \, \right] \, \chi(x) \;,
\end{equation}
where $m_{\chi }$ is the fermion mass, and clearly $\chi $ is a fermion
field of $N^{2}$ components. In accord with the symmetry $SU(N)\times SU(N)$%
, the field $\chi $ can be split into the components
\begin{equation}\label{Qiqi}
q_{i} \hspace{0.3cm} \mbox{with} \hspace{0.3cm}
i=1,2,...,\frac{N(N-1)}{2}
\hspace{0.5cm}\mbox{and}\hspace{0.5cm}
Q_{i}
\hspace{0.3cm} \mbox{with} \hspace{0.3cm}
i=1,2,...,\frac{N(N+1)}{2} \; ,
\end{equation}
where $N\times N=\frac{N(N-1)}{2}\oplus \frac{N(N+1)}{2}$. The components
of $q$ and $Q$ have the following transformations
\begin{equation}\label{transfqQ}
q_{i}\longmapsto q_{i}^{\;\prime }=(e^{ \, i \, \omega^{a} \, t^{a}})_{ij} \, q_{j}
\hspace{0.5cm}\mbox{and}\hspace{0.5cm}
Q_{i}\longmapsto Q_{i}^{\;\prime }=(e^{ \, i \, \omega
^{a} \, \Lambda ^{a}})_{ij} \, Q_{j} \; ,
\end{equation}
in which $t^{a}$ and $\Lambda ^{a}$ are square matrices $\frac{N(N-1)}{2}%
\times \frac{N(N-1)}{2}$ and $\frac{N(N+1)}{2}\times \frac{N(N+1)}{2}$, both
in the fundamental representation of $SU(N)$. Now one wishes introducing
different masses for the two sets of fermions $q$ and $Q$, but the massive
term of (\ref{Lfermions}) is incompatible with this requirement because the
field $\chi $ is a mixing of $q$ and $Q$. Consequently, the set of
transformations (\ref{transfqQ}) and (\ref{transfglobais12}) form our
fermionic sector of $SU(N)\times SU(N)$, so we propose the lagrangian
invariant by those transformations
\begin{equation}\label{LfermionsqQ}
\mathcal{L}_{fermions}=\bar{\chi}(x) \, i \, \gamma ^{\mu} \, D_{\mu }(G,C) \, \chi(x) -m_{q} \,
\bar{q} \, q-m_{Q} \, \bar{Q} \, Q \; ,
\end{equation}
where $m_{q}$ and $m_{Q}$ are masses of $q$ and $Q$, respectively.
%
%
%
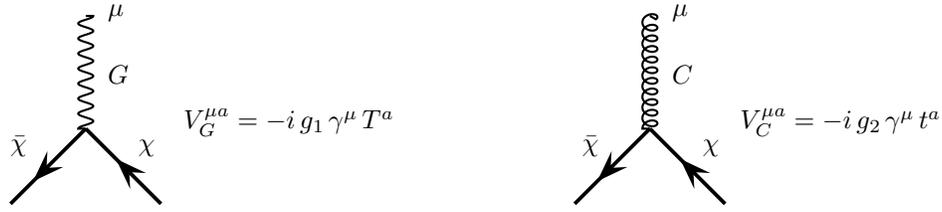
\begin{figure}[h]
\begin{center}
\newpsobject{showgrid}{psgrid}{subgriddiv=1,griddots=10,gridlabels=6pt}
\begin{pspicture}(1,0)(9,1.5)
\psset{arrowsize=0.2 2}
%
%
\pscoil[coilarm=0,coilaspect=0,coilwidth=0.2,coilheight=1.0,linecolor=black](0,0)(0,1.5)
\psline[linecolor=black,linewidth=0.5mm]{->}(1,-1)(0.4,-0.4)
\psline[linecolor=black,linewidth=0.5mm](0.5,-0.5)(0,0)
\psline[linecolor=black,linewidth=0.5mm]{->}(0,0)(-0.7,-0.7)
\psline[linecolor=black,linewidth=0.5mm](-0.5,-0.5)(-1,-1)
\put(0.3,1.5){$\mu$}
\put(0.3,0.6){$G$}
\put(-1,-0.3){$\bar{\chi}$}
\put(0.7,-0.3){$\chi$}
\put(1.3,0){$V_{G}^{\mu a}=-i \, g_{1} \, \gamma^{\mu} \, T^{a}$}
%
%
%
%
\pscoil[coilarm=0,coilwidth=0.2,coilheight=1.0,linecolor=black](7.5,0)(7.5,1.5)
\psline[linecolor=black,linewidth=0.5mm]{->}(8.5,-1)(7.9,-0.4)
\psline[linecolor=black,linewidth=0.5mm](8,-0.5)(7.5,0)
\psline[linecolor=black,linewidth=0.5mm]{->}(7.5,0)(6.8,-0.7)
\psline[linecolor=black,linewidth=0.5mm](7,-0.5)(6.5,-1)
\put(7.8,1.5){$\mu$}
\put(7.8,0.6){$C$}
\put(6.6,-0.3){$\bar{\chi}$}
\put(8.2,-0.3){$\chi$}
\put(8.7,0){$V_{C}^{\mu a}=-i \, g_{2} \, \gamma^{\mu} \, t^{a}$}
\end{pspicture}
\vspace{1.2cm}
\end{center}
\caption{Fermions vertices interacting with massless and massive vectorials
fields.}
\label{IntChiGC}
\end{figure}
%

Now we will apply it to the particular case $SU_{c}(3)\times SU_{c}(3)$
which can be interesting due to preserve the QCD color symmetry. The vector
fields sector is like those showed earlier, but now we have eight massless
gauge field plus eight massive vector fields $\left( \, G^{\mu \, a} \, , \, C^{\mu \, a} \, \right)$
with $a=\{1,2,...,8\}$. Both can be interpreted as eight massless gluons plus
eight massive gluons. The sixteen matrices $\{T^{a}\}$ and $\{t^{a}\}$ $%
(a=1,2,...8)$ build up the basis for gluons massless and massive gluons,
respectively, being the combinations (\ref{Tt}) of the Gell-Mann matrices
\cite{MutaPokoriskiLivro} of group $SU(3)$ in the fundamental representation
\begin{equation}\label{TtSU3}
T^{a}=\frac{\lambda ^{a}}{2}\otimes \mathbf{1}+\mathbf{1}\otimes \frac{%
\lambda ^{a}}{2}\hspace{0.5cm}\mbox{and}\hspace{0.5cm}t^{a}=\frac{\lambda
^{a}}{2}\otimes \mathbf{1}-\mathbf{1}\otimes \frac{\lambda ^{a}}{2}\;,
\end{equation}
satisfying the commutation relation (\ref{RelComutTt}), and the
structure constant of group $f^{abc}$ listed by table in \cite{MutaPokoriskiLivro}.


The sector of the composite quarks is defined by the direct product
\cite{DoriaHelayelMario}
\begin{eqnarray}\label{CampoChiQuarks}
\chi = \left(
\begin{array}{c}
\psi_{1} \\
\psi_{2} \\
\psi_{3} \\
\end{array}
\right) \otimes \left(
\begin{array}{c}
\phi_{1} \\
\phi_{2} \\
\phi_{3} \\
\end{array}
\right) \; ,
\end{eqnarray}
in which the quark is a composition of fermions $\psi=\left( \psi_{1},\psi_{2},
\psi_{3} \right)$ and scalars $\phi=\left( \phi_{1},\phi_{2},\phi_{3} \right)$, of three colors
each, both in the fundamental representation. Clearly, these product is a
column matrix of nine components. It can be split in quarks triplets
\begin{eqnarray}\label{quarkstripletos}
q_{i}=\frac{1}{\sqrt{2}} \, \epsilon_{ijk} \, \psi_{j} \, \phi_{k} \; ,
\hspace{0.5cm} \mbox{with} \hspace{0.5cm} i,j,k = 1,2,3 \; ,
\end{eqnarray}
where $\epsilon_{ijk}$ is Levi-Civita symbol, and sextets quarks are defined
by
\begin{equation}\label{quarkssextetos}
\Xi_{(ij)}=\frac{1}{\sqrt{2}}\left(\psi_{i} \, \phi_{j}+\psi_{j} \, \phi_{i}\right)
\; , \hspace{0.5cm} \mbox{with} \hspace{0.5cm} i,j = 1,2,3 \; ,
\end{equation}
in accord with the components
\begin{eqnarray}\label{componentesquarkssextetos}
\sqrt{2} \, Q_{1}:=\Xi_{(11)}=\sqrt{2} \, \psi_{1} \, \phi_{1}
\hspace{0.3cm} , \hspace{0.3cm} \sqrt{2} \, Q_{4}:=\Xi_{(22)}=\sqrt{2} \, \psi_{2} \, \phi_{2}
\hspace{0.3cm} , \hspace{0.3cm}
\sqrt{2} \, Q_{6}:=\Xi_{(33)}=\sqrt{2} \, \psi_{3} \, \phi_{3}  \notag \\
Q_{2}:=\Xi_{(12)}=\frac{1}{\sqrt{2}} \, \left(\psi_{1} \, \phi_{2}+\psi_{2} \, \phi_{1}
\right) \hspace{0.3cm} , \hspace{0.3cm} Q_{3}:=\Xi_{(13)}=\frac{1}{\sqrt{2}}
\left(\psi_{1} \, \phi_{3}+\psi_{3} \, \phi_{1}\right)  \notag \\
Q_{5}:=\Xi_{(23)}=\frac{1}{\sqrt{2}}\left(\psi_{2} \, \phi_{3}+\psi_{3} \, \phi_{2}
\right) \; , \hspace{3.5cm}
\end{eqnarray}
then, one obtains
\begin{eqnarray}\label{matriztripletossextetos}
\chi=\left(
\begin{array}{c}
Q_{1} \\
\frac{1}{\sqrt{2}} \, (q_{3}+Q_{2}) \\
\frac{1}{\sqrt{2}} \, (Q_{3}-q_{2}) \\
\frac{1}{\sqrt{2}} \, (Q_{2}-q_{3}) \\
Q_{4} \\
\frac{1}{\sqrt{2}} \, \left(q_{1}+Q_{5}\right) \\
\frac{1}{\sqrt{2}} \, (Q_{3}+q_{2}) \\
\frac{1}{\sqrt{2}} \, (Q_{5}-q_{1}) \\
Q_{6} \\
\end{array}
\right) \; .
\end{eqnarray}
Thus, introducing the definition where $q$ and $Q$ are respectively the
column matrices of triplets and sextets quarks
\begin{eqnarray}\label{qQ}
q=\left(
\begin{array}{c}
q_{1} \\
q_{2} \\
q_{3} \\
\end{array}
\right) \hspace{0.5cm} \mbox{and} \hspace{0.5cm} Q=\left(
\begin{array}{c}
Q_{1} \\
Q_{2} \\
Q_{3} \\
Q_{4} \\
Q_{5} \\
Q_{6} \\
\end{array}
\right) \; ,
\end{eqnarray}
one gets,
\begin{eqnarray}\label{LQTripSext}
\mathcal{L}_{quarks}=\mathcal{L}_{quarks-0}+\mathcal{L}_{quarks-int} \; ,
\end{eqnarray}
where
\begin{eqnarray}\label{LQFree}
\mathcal{L}_{quarks-0}=\bar{q} \, \left( \, i \, \gamma^{\mu} \, \partial_{\mu}-m_{q} \, {\bf 1} \, \right) \, q
+\bar{Q}\left( \, i \, \gamma^{\mu} \, \partial_{\mu}-m_{Q} \, {\bf 1} \, \right) Q \; ,
\end{eqnarray}
in which $m_{q}$ and $m_{Q}$ set the masses of triplets and sextets,
respectively. The triplets and sextets have symmetry transformations given
by (\ref{transfqQ}), in which $t^{a}=\lambda^{a}/2 \; \left( \, a=1,2,...,8 \, \right)$ are
Gell-Mann matrices, and $\Lambda^{a} \; \left( \, a=1,2,...,8 \, \right)$ are square matrices $%
6 \times 6$ listed in \cite{DoriaHelayelMario}. Intuitively, one expects
massive sextets quarks will appear at higher energies than usual $QCD$. %
%
%
\begin{figure}[!h]
\begin{center}
\newpsobject{showgrid}{psgrid}{subgriddiv=1,griddots=10,gridlabels=6pt}
\begin{pspicture}(0,0)(4.5,2.8)
%
\psline[linewidth=0.1,linecolor=black]{->}(5.5,2)(6.8,2)
\psline[linewidth=0.1,linecolor=black](6.5,2)(7.5,2)
%
%
\large{\put(-3.7,2){$\langle \, \bar{q}_{i} \, q_{j} \, \rangle
=\frac{i \, \delta_{ij}}{\slash\!\!\!p-m_{q}+i \, \varepsilon}$}}
\large{\put(0.5,2){$\langle \, \bar{Q}_{i} \, Q_{j} \, \rangle
=\frac{i \, \delta_{ij}}{\slash\!\!\!p-m_{Q}+i \, \varepsilon}$}}
%
%
\put(5.4,2.2){$i$}
\put(7.4,2.2){$j$}
\put(6.45,1.5){$p$}
%
%
%
\end{pspicture}
\vspace{-1.3cm}
\end{center}
\caption{Quarks propagators.}
\label{propagatorsqQ}
\end{figure}

The interactions terms involving the triplets, sextets quarks and gluons are
\begin{eqnarray}  \label{LQuarksInt}
\mathcal{L}_{int-qQG}^{I}=-g_{1} \, \bar{\chi}_{i} \, G_{\mu}^{a} \, \left(T^{a}\right)_{ij}
\chi_{j} =-\frac{1}{2} \, g_{1} \, \bar{q} \, \gamma^{\mu} \, \left(
\begin{array}{ccc}
G_{\mu}^{3}+\frac{G_{\mu}^{8}}{\sqrt{3}} & G_{\mu}^{1}-iG_{\mu}^{2} &
G_{\mu}^{4}-iG_{\mu}^{5} \\
G_{\mu}^{1}+iG_{\mu}^{2} & G_{\mu}^{3}-\frac{G_{\mu}^{8}}{\sqrt{3}} &
G_{\mu}^{6}-iG_{\mu}^{7} \\
G_{\mu}^{4}+iG_{\mu}^{5} & G_{\mu}^{6}+iG_{\mu}^{7} & -\frac{2G_{\mu}^{8}}{%
\sqrt{3}} \\
&  &
\end{array}
\right)q \hspace{2cm}  \notag \\
-\frac{1}{2} \, g_{1} \, \bar{Q} \, \gamma^{\mu} \, \left(
\begin{array}{cccccc}
G_{\mu}^{3}+\frac{G_{\mu}^{8}}{\sqrt{3}} & G_{\mu}^{1}+iG_{\mu}^{2} &
G_{\mu}^{4}+iG_{\mu}^{5} & G_{\mu}^{4}+iG_{\mu}^{5} & 0 & 0 \\
G_{\mu}^{1}-iG_{\mu}^{2} & \frac{2G_{\mu}^{8}}{\sqrt{3}} &
G_{\mu}^{6}+iG_{\mu}^{7} & 0 & G_{\mu}^{4}+iG_{\mu}^{5} & 0 \\
G_{\mu}^{4}-iG_{\mu}^{5} & G_{\mu}^{6}-iG_{\mu}^{7} & G_{\mu}^{3}-\frac{%
G_{\mu}^{8}}{\sqrt{3}} & 0 & G_{\mu}^{1}+iG_{\mu}^{2} & G_{\mu}^{4}+iG_{%
\mu}^{5} \\
0 & G_{\mu}^{1}-iG_{\mu}^{2} & 0 & G_{\mu}^{3}-\frac{G_{\mu}^{8}}{\sqrt{3}}
& G_{\mu}^{6}+iG_{\mu}^{7} & 0 \\
0 & G_{\mu}^{4}-iG_{\mu}^{5} & G_{\mu}^{1}-iG_{\mu}^{2} &
G_{\mu}^{6}-iG_{\mu}^{7} & -G_{\mu}^{3}-\frac{G_{\mu}^{8}}{\sqrt{3}} &
G_{\mu}^{6}+iG_{\mu}^{7} \\
0 & 0 & G_{\mu}^{4}-iG_{\mu}^{5} & 0 & G_{\mu}^{6}-iG_{\mu}^{7} & -\frac{%
2G_{\mu}^{8}}{\sqrt{3}} \\
&  &  &  &  &
\end{array}
\right) \, Q \; ,
\end{eqnarray}
and
\begin{eqnarray}\label{LintQq}
\mathcal{L}_{int-qQC}=-g_{2} \, \bar{\chi}_{i} \, C_{\mu}^{a} \, \left(t^{a}\right)_{ij}
 \chi_{j}= -\frac{1}{2} \, g_{2} \, \bar{Q} \, \gamma^{\mu} \, \left(
\begin{array}{ccc}
0 & -C_{\mu}^{4}-iC_{\mu}^{5} & -C_{\mu}^{1}-iC_{\mu}^{2} \\
-C_{\mu}^{4}-iC_{\mu}^{5} & C_{\mu}^{6}+iC_{\mu}^{7} & 0 \\
C_{\mu}^{1}+iC_{\mu}^{2} & 0 & -C_{\mu}^{6}+iC_{\mu}^{7} \\
-C_{\mu}^{6}-iC_{\mu}^{7} & 0 & C_{\mu}^{1}-iC_{\mu}^{2} \\
0 & 0 & C_{\mu}^{4}-iC_{\mu}^{5} \\
C_{\mu}^{6}-iC_{\mu}^{7} & -C_{\mu}^{4}+iC_{\mu}^{5} & 0 \\
&  &
\end{array}
\right) \, q \hspace{2cm}  \notag \\
-\frac{1}{2} \, g_{2} \, \bar{q} \, \gamma^{\mu} \, \left(
\begin{array}{cccccc}
0 & -C_{\mu}^{4}+iC_{\mu}^{5} & 0 & -C_{\mu}^{6}+iC_{\mu}^{7} & 0 &
C_{\mu}^{6}+iC_{\mu}^{7} \\
C_{\mu}^{4}-iC_{\mu}^{5} & C_{\mu}^{6}-iC_{\mu}^{7} & -C_{\mu}^{3}-\sqrt{3}%
C_{\mu}^{8} & 0 & -C_{\mu}^{1}-iC_{\mu}^{2} & -C_{\mu}^{4}-iC_{\mu}^{5} \\
-C_{\mu}^{1}+iC_{\mu}^{2} & 0 & 0 & C_{\mu}^{1}+iC_{\mu}^{2} &
C_{\mu}^{4}+iC_{\mu}^{5} & 0 \\
&  &  &  &  &
\end{array}
\right) \, Q \; .
\end{eqnarray}
%
%
%
%
\begin{figure}[!h]
\begin{center}
\newpsobject{showgrid}{psgrid}{subgriddiv=1,griddots=10,gridlabels=6pt}
\begin{pspicture}(1,0)(11,1.6)
\psset{arrowsize=0.2 2}
%
%
\pscoil[coilarm=0,coilaspect=0,coilwidth=0.2,coilheight=1.0,linecolor=black](0,0)(0,1.5)
\psline[linecolor=black,linewidth=0.5mm]{->}(1,-1)(0.4,-0.4)
\psline[linecolor=black,linewidth=0.5mm](0.5,-0.5)(0,0)
\psline[linecolor=black,linewidth=0.5mm]{->}(0,0)(-0.7,-0.7)
\psline[linecolor=black,linewidth=0.5mm](-0.5,-0.5)(-1,-1)
\put(0.3,1.4){$\mu$}
\put(0.3,0.6){$G$}
\put(-1,-0.3){$\bar{q}$}
\put(0.7,-0.3){$q$}
\put(1.3,0){$V_{G \; ij}^{\mu a}=-i \, g_{1} \, \gamma^{\mu} \, \left(T^{a}\right)_{ij}$}
%
%
%
\pscoil[coilarm=0,coilaspect=0,coilwidth=0.2,coilheight=1.0,linecolor=black](7.5,0)(7.5,1.5)
\psline[linecolor=black,linewidth=0.5mm]{->}(8.5,-1)(7.9,-0.4)
\psline[linecolor=black,linewidth=0.5mm](8,-0.5)(7.5,0)
\psline[linecolor=black,linewidth=0.5mm]{->}(7.5,0)(6.8,-0.7)
\psline[linecolor=black,linewidth=0.5mm](7,-0.5)(6.5,-1)
\put(7.8,1.4){$\mu$}
\put(7.8,0.6){$G$}
\put(6.5,-0.3){$\bar{Q}$}
\put(8.2,-0.3){$Q$}
\put(8.8,0){$V_{G\;ij}^{\mu a}=-i \, g_{1} \, \gamma^{\mu} \, \left(T^{a}\right)_{ij}$}
\end{pspicture}
\vspace{1.2cm}
\end{center}
\caption{Vertices of triplets and sextets Quarks interacting with massless
gluons.}
\label{QTSGluons}
\end{figure}
%
%
%
%
%
\begin{figure}[!h]
\begin{center}
\newpsobject{showgrid}{psgrid}{subgriddiv=1,griddots=10,gridlabels=6pt}
\begin{pspicture}(1,0)(11,1.5)
\psset{arrowsize=0.2 2}
%
%
\pscoil[coilarm=0,coilwidth=0.2,coilheight=1.0,linecolor=black](0,0)(0,1.5)
\psline[linecolor=black,linewidth=0.5mm]{->}(1,-1)(0.4,-0.4)
\psline[linecolor=black,linewidth=0.5mm](0.5,-0.5)(0,0)
\psline[linecolor=black,linewidth=0.5mm]{->}(0,0)(-0.7,-0.7)
\psline[linecolor=black,linewidth=0.5mm](-0.5,-0.5)(-1,-1)
\put(0.3,1.4){$\mu$}
\put(0.3,0.6){$C$}
\put(-1,-0.3){$\bar{Q}$}
\put(0.7,-0.3){$q$}
\put(1.3,0){$V_{C \; ij}^{\mu a}=-ig_{2}\gamma^{\mu}\left(t^{a}\right)_{ij}$}
%
%
%
\pscoil[coilarm=0,coilwidth=0.2,coilheight=1.0,linecolor=black](7.5,0)(7.5,1.5)
\psline[linecolor=black,linewidth=0.5mm]{->}(8.5,-1)(7.9,-0.4)
\psline[linecolor=black,linewidth=0.5mm](8,-0.5)(7.5,0)
\psline[linecolor=black,linewidth=0.5mm]{->}(7.5,0)(6.8,-0.7)
\psline[linecolor=black,linewidth=0.5mm](7,-0.5)(6.5,-1)
\put(7.8,1.4){$\mu$}
\put(7.8,0.6){$C$}
\put(6.5,-0.3){$\bar{q}$}
\put(8.2,-0.3){$Q$}
\put(8.8,0){$V_{C\;ij}^{\mu a}=-ig_{2}\gamma^{\mu}\left(t^{a}\right)_{ij}$}
\end{pspicture}
\vspace{1.2cm}
\end{center}
\caption{Vertices of triplets and sextets Quarks interacting with massive
gluons.}
\label{QTSGluons}
\end{figure}
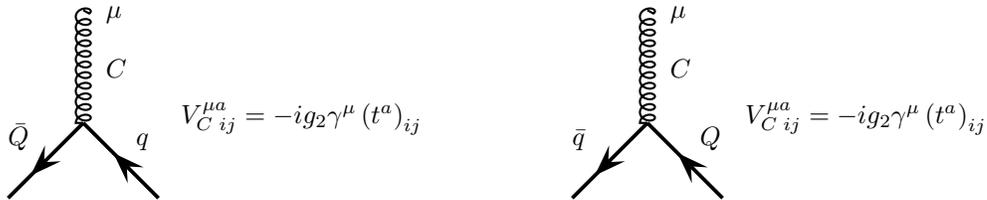
%
%


The constants coupling $g_{1}$ and $g_{2}$ sets the interaction between
massless gluons and massive gluons with quarks, respectively. These terms
show the interaction between triplets quarks intermediates by massless
gluons like in usual $QCD$, furthermore the sextets quarks also interact by
means of massless gluons. In the case of massive gluons, they appear just as
mediators of the interactions between triplet and sextets of quarks.

For end of this section we resume what one has obtained here. Thus we have
got a complete effective quantum lagrangian for symmetry $SU(N)\times SU(N)$
governed by transformations (\ref{transfgaugeGCSep}), where vector fields,
Faddeev-Popov and quarks sectors are given by
\begin{equation}\label{Lterms}
\mathcal{L}_{eff}=\mathcal{L}_{vector\;fields}+\mathcal{L}_{FP}+\mathcal{L}%
_{gauge\;fixing}+\mathcal{L}_{quarks}\;.
\end{equation}
%
%
%
These composition of fields is a way to introduce sextets of quarks beyond
already known triplets of quarks in the particular case $SU_{c}(3)\times
SU_{c}(3)$. For case $SU_{c}(3)\times SU_{c}(3)$, one gets eight massless
gluons plus eight massive gluons by self-interacting and with triplets and
sextets of quarks. Curiously, anti-triplets and sextets (or anti-sextets and
triplets) of quarks interact by means of massive gluons only, while the
massless gluons are just mediators of the interaction of anti-triplets with
triplets, or anti-sextets with sextets. %
%

\section{\textsc{Remarks on renormalization}}
In this section one begins an essential program to establish the full
renormalization of the model $SU(N) \times SU(N)$. We have obtained the
complete lagrangian invariant by transformations (\ref{transfgaugeGCSep}),
and here, we will analyze the corresponding power counting.
The structure of the radiative corrections 
has Feynman integrals divergent that behave like those in the Yang-Mills
symmetry.

\subsection{Power counting}

The analysis of power counting is useful to indicate us on behavior of all
possible Feynman diagrams that contribute to the model in higher order in
the perturbation series. Based on structure of propagators and vertices have
been showed earlier, we shall obtain the superficial divergence degree $D$
of any Feynman diagrams that those symmetry permits. Hence one defines the
following notation for external and internal lines, vertices, loops of
massless, massive gluons, quarks and ghosts :
\begin{eqnarray}  \label{DefinitionsLVIE}
D\!\!\!&=&\!\!\!\mbox{Superficial divergence degree for any Feynman diagram}
\notag \\
2\omega\!\!\!&=&\!\!\!\mbox{Space-time dimension}  \notag \\
L\!\!\!&=&\!\!\!\mbox{Number of loops}  \notag \\
I_{G}\!\!\!&=&\!\!\!\mbox{Number of internal lines of massless gluons}
\notag \\
I_{C}\!\!\!&=&\!\!\!\mbox{Number of internal lines of massive gluons}  \notag
\\
I_{g}\!\!\!&=&\!\!\!\mbox{Number of internal lines of ghosts}  \notag \\
I_{\chi}\!\!\!&=&\!\!\!\mbox{Number of internal lines of quarks}  \notag \\
V_{3G}\!\!\!&=&\!\!\!\mbox{Three line vertex of massless gluons}  \notag \\
V_{4G}\!\!\!&=&\!\!\!\mbox{Four line vertex of massless gluons}  \notag \\
V_{3GC}\!\!\!&=&\!\!\!%
\mbox{Three line vertex that mix massless and massive
gluons}  \notag \\
V_{4GC}\!\!\!&=&\!\!\!%
\mbox{Four line vertex that mix massless and massive
gluons}  \notag \\
V_{4C}\!\!\!&=&\!\!\!\mbox{Four line vertex of massive gluons}  \notag \\
V_{g}\!\!\!&=&\!\!\!\mbox{Vertex of ghost interaction}  \notag \\
V_{\chi G}\!\!\!&=&\!\!\!%
\mbox{Vertex of quarks and massless gluons
interaction}  \notag \\
V_{\chi C}\!\!\!&=&\!\!\!%
\mbox{Vertex of quarks and massive gluons
interaction}  \notag \\
E_{G}\!\!\!&=&\!\!\!\mbox{Number of external lines of massless gluons}
\notag \\
E_{C}\!\!\!&=&\!\!\!\mbox{Number of external lines of massive gluons}  \notag
\\
E_{\chi}\!\!\!&=&\!\!\!\mbox{Number of external lines of quarks} \; .
\end{eqnarray}
With all those definitions, the divergence degree of any Feynman graphic has
the expression
\begin{eqnarray}  \label{degreeD}
D=2 \, \omega \, L-2I_{G}-2I_{C}-2I_{g}-I_{\chi}+V_{3G}+V_{3GC}+V_{g} \; ,
\end{eqnarray}
the number of loops is
\begin{eqnarray}  \label{Lloops}
L=I_{G}+I_{C}+I_{g}+I_{\chi}-V_{3G}-V_{4G}-V_{3GC}-V_{4GC}-V_{4C}-V_{g}-V_{%
\chi G}-V_{\chi C}+1 \; ,
\end{eqnarray}
and the topological relations
\begin{eqnarray}  \label{Toprelations}
2I_{G}+E_{G}\!\!\!&=&\!\!\!4V_{4G}+3V_{3G}+2V_{4GC}+V_{3GC}+V_{g} \; ,
\notag \\
2I_{C}+E_{C}\!\!\!&=&\!\!\!2V_{3GC}+2V_{4GC}+4V_{4C} \; ,  \notag \\
2I_{\chi}+E_{\chi} \!\!\!&=&\!\!\! 2V_{\chi G}+2V_{\chi C} \; ,
\end{eqnarray}
By substituting (\ref{Lloops}) and (\ref{Toprelations}) in (\ref{degreeD}),
we find the divergence degree $D$ in terms of external lines and vertices
\begin{equation}  \label{degreeDEV}
D=2\omega+(1-\omega)(E_{G}+E_{C})+\left(\frac{1}{2}-\omega\right)E_{\chi}+(2
\omega-4)\left(V_{4G}+\frac{1}{2} \, V_{3G} +\frac{1}{2} \, V_{g}
+\frac{1}{2} \, V_{3GC}+V_{4GC}+V_{4C}\right) \; ,
\end{equation}
where we have used that there is no any external ghost, so $I_{g}=V_{g}$.
These result give us a important interpretation in the case of the physical
dimension $\omega=2$. In this case the expression (\ref{degreeDEV}) depends
on external lines of massless and massive gluons only
\begin{eqnarray}  \label{degreeDEV4D}
D=4-E_{G}-E_{C}-\frac{3}{2} \, E_{\chi} \; ,
\end{eqnarray}
The equation (\ref{degreeDEV4D}) is a great indication that this model is
renormalizable in four-dimension, like in the usual Yang-Mills symmetry. The
dimensionality analysis for constants coupling $\left( \, g_{1},g_{2},g_{3} \, \right)$ is
totally analogous to the $SU(N)$ case. Consider the generator functional
\begin{eqnarray}  \label{Z}
Z \sim \int \mathcal{D}G\mathcal{D}C\mathcal{D}\bar{\eta}\mathcal{D}\eta \;
\exp \left[ i\int d^{2\omega}x \; \mathcal{L}_{total} \right] \; ,
\end{eqnarray}
where we have substituted the physical dimension by the dimensional
regularizator. Clearly if the action is dimensionless, the fields dimensions
are given by
\begin{eqnarray}  \label{DimensionsGCg}
\left[G\right]=\left[C\right]=\left[\eta\right]=[\Lambda]^{\omega-1} \; ,
\end{eqnarray}
in which $\left[\Lambda\right]$ sets a mass dimension parameter. By using
those relations in the interactions terms of (\ref{LInt3}) and (\ref{LKInt4})
one gets the dimension of coupling constants
\begin{eqnarray}  \label{Dimensionsg}
[g_{1}]=[g_{2}]=[g_{3}]=[\Lambda]^{2-\omega} \; .
\end{eqnarray}
These relations shows that in the case of physical dimension $\omega=2$ all
coupling constants of this model are dimensionless, it is another statement
that establish the renormalizability of the model. In the next subsection
one shall present the perturbative character of the model by writing all
contributions to one-loop for all propagators and vertex. These
contributions are clearly divergent, but their divergent structure has a
behavior controllable like in the usual Yang-Mills case.

\subsection{Conclusions}
We have studied a Yang-Mills extension based on composition of two
independents non-abelian groups $SU(N)$. This symmetry is constructed in
such a way that fermions are composite of a direct product between others
fermions and scalars. It yields a gauge sector where fields follow (\ref%
{transfgaugeGCSep}) transformations. The first one is just gauge
transformation of a non-abelian field $G_{\mu }$ Lie algebra valued in a
given basis, while the second is just an unitary massive vector field $%
C_{\mu }$ Lie algebra valued in a second basis. Under this deduction, one
intends to go beyond $QCD$ through lagrangian (\ref{Lterms}).
It introduces new possibilities beyond those already known from usual
Yang-Mills symmetry.

The first effort of this work is to show the lagrangians (\ref%
{lagrangianoZmunu}) and (\ref{Lfermions}) validity for perturbation theory.
In a previous work one has proved on its hamiltonian positivity \cite%
{DoriaHelayelMario}. Given such stability for perturbative approach a next
step should be to study on its unitarity and renormalizability. The
unitarity at the tree level of model is satisfied by establishing conditions
between the parameters $(a,c)$ that set the transversal and longitudinal
parts, in accord with positivity of residue into the propagator of $C_{\mu}$%
. The equation (\ref{degreeDEV4D}) indicates renormalizability in terms of
power counting. Thus one gets a health model in terms of perturbation
theory. Its hamiltonian is not negative, the correspondent power counting
analysis supports renormalizability, unitarity is satisfied at tree level
and free of anomalies. Consequently, it is a model candidate for being
studied under Callan-Symmanzik equation.



The full renormalizability of the model is something to be studied in the
next paper. Here we have presented the BRST symmetry and Slavnov-Taylor
identities as a beginning way, in which one shall enable to establish
relationship between the one-particle irreducible Green functions. Thus as a
next effort we will calculate the Feynman diagrams to one loop approximation
and so to realize a study on renormalization group and the Callan-Symanzik
beta function for $SU(N)\times SU(N)$. Considering the trilinear vertices
abundance one expects that in this case it be more asymptotically free than
those $QCD$ usual case. The addition of composite scalars quarks on behavior
of beta function must also analyzed, that is, if this model is
asymptotically free in the presence of scalars quarks under a lagrangian $%
(D_{\mu}\Phi_{i})^2$ where $\Phi_{i}$ scalar field is constituted by scalar
colorful stones $\Phi_{i}=f_{ijk} \, \phi_{i} \, \phi_{j}$ \cite{DoriaHelayelMario}.
The calculus of deep inelastic scattering of $SU_{c}(3)\times SU_{c}(3)$ is
something to be investigated and the influence due to the presence of scalar
quarks and massive gluons on results.

Thus a new color phenomenology is proposed for $LHC$. Based on $SU_{c}(3)$
symmetry there is new suggestions for the colorful world. The model $%
SU_{c}(3)\times SU_{c}(3)$ or double $SU_{c}(3)$ contains $QCD$ and extends
it for composite quarks (fermionic and bosonic) in triplets and sextets,
massless and massive gluons (transverse and longitudinal) as another
possibilities derived from the Twelve colorful stones table in \cite%
{DoriaHelayelMario}. It provides a dynamics for quarks and leptons with
different spins. In terms of interactions, one expects a weaker running
coupling constant than $QCD$ due to a larger number of three and four gluons
vertices. This means that there is a colorful weak interaction for being
investigated theoretically. It is a model where one replaces colorful
massive gluons instead of $(W^{\pm},Z^{0})$ as the intermediate bosons for
flavors exchange.


Thus, given $LHC$ new energy range, one expects new experimental
possibilities for the colorful world. Instead of just following the pattern
with quarks in triplets and eight gluons there is a new colorful diversity
to be measured from an origin based on twelve colorful stones coupled to a
double symmetry $SU_{c}(3)$. As a new phenomenological sector, one expects
massive gluons, scalar quarks, quarks in sextets with different masses than
the usual ones, massive glueballs and also a new variety of exotic mesons
and barions as $\Theta^{+}$ to be detected by $LHC$ \cite%
{NakanoBarminStepanyan}.



\end{document}